\documentclass[12pt]{article}
\usepackage{lineno}
\usepackage[colorlinks, linkcolor=black, citecolor=black]{hyperref}
\usepackage{epsfig,epsf,psfrag,multirow,rotating}
\usepackage[latin1]{inputenc}
\usepackage{enumerate,enumitem,tikz}
\usepackage{lscape}
\usepackage{changepage,bbm,comment}
\usepackage{color,latexsym,graphicx}
\usepackage{amsfonts,amssymb,amscd,amsmath,amsbsy}
\usepackage{xcolor}
\usepackage{natbib}

\usepackage{mathtools}

\setlist[enumerate]{
  labelindent=0pt,
  leftmargin=*,
}

\setlist[itemize]{
  labelindent=0pt,
  leftmargin=*,
}

\usepackage{caption}
\usepackage{setspace} 

\usepackage{sectsty}
\sectionfont{\fontsize{12}{15}\selectfont}
\subsectionfont{\fontsize{12}{15}\selectfont}

\usepackage{hyperref}

\usepackage{amsfonts, amsmath}
\usepackage{graphicx, epstopdf}
\usepackage{color,xcolor}
\usepackage{float, multirow, booktabs}

\newcommand{\bas}{\begin{eqnarray*}}
\newcommand{\eas}{\end{eqnarray*}}
\newcommand{\ba}{\begin{eqnarray}}
\newcommand{\ea}{\end{eqnarray}}
\newcommand{\bit}{\begin{itemize}}
\newcommand{\eit}{\end{itemize}}
\newcommand{\pr}{{\rm pr}}
\newcommand{\e}{ { \mathbb{E}}}

\newcommand{\var}{{\mathbb{V}\rm ar}}

\def\T{{ \mathrm{\scriptscriptstyle \top} }}
\newtheorem{assumption}{Assumption}

\newtheorem{theorem}{Theorem}

\newtheorem{example}{Example}
\newtheorem{remark}{Remark}

\def\no{\noindent}

\definecolor{darkblue}{rgb}{0.0, 0.0, 0.5}

\mathtoolsset{showonlyrefs=true}

\begin{document}

{\centering {\large {\bf Instability of inverse probability weighting methods and a remedy for non-ignorable missing data}} \par}

\bigskip

\centerline{
Pengfei Li,
Jing Qin,
and Yukun Liu 
}

\bigskip

\bigskip

\hrule

{\small
\begin{quotation}
\no
Inverse probability weighting (IPW) methods are commonly used to analyze non-ignorable missing data under the assumption of a logistic model for the missingness probability. However, solving IPW equations numerically may involve non-convergence problems when the sample size is moderate and the missingness probability is high. Moreover, those equations often have multiple roots, and identifying the best root is challenging. Therefore, IPW methods may have low efficiency or even produce biased results. We identify the pitfall in these methods pathologically: they involve the estimation of a moment-generating function, and such functions are  notoriously unstable in general. As a remedy, we model the outcome distribution given the covariates of the completely observed individuals semiparametrically. After forming an induced logistic regression model for the missingness status of the outcome and covariate, we develop a maximum conditional likelihood method to estimate the underlying parameters. The proposed method circumvents the estimation of a moment-generating function and hence overcomes the instability of IPW methods. Our theoretical and simulation results show that the proposed method outperforms existing competitors greatly. Two real data examples are analyzed to illustrate the advantages of our method. We conclude that if only a parametric logistic regression is assumed but the outcome regression model is left arbitrary, then one has to be cautious in using any of the existing statistical methods in problems involving non-ignorable missing data.

\vspace{0.3cm}

\no
\end{quotation}
}

\hrule

\bigskip

\bigskip

\section{Introduction}
\label{s:intro}

Problems involving non-ignorable missing data (NIMD) have attracted much attention in recent years \citep{Tang2018} because they are encountered frequently in many areas but are much more challenging to handle than are problems involving missing at random (MAR) data. Data are MAR if the missingness probability depends on only the observed data and not the unobserved data; otherwise they are non-ignorable missing or missing not at random \citep{Little2002, Rubin1987}. We assume that the outcome variable $Y$ may be missing and that the covariate $X$ is always observable. Denote the missingness indicator of $Y$ as $R$, which equals $0$ if $Y$ is missing or $1$ otherwise.

Two facts make NIMD more challenging to handle than MAR data. First, for MAR data, the (conditional) missingness probability or propensity score $\pr(R=1|X, Y) = \pr(R=1|X)$ and the (conditional) outcome regression function $\pr(Y|X)$ are separable in the likelihood, therefore inferences can be made for either of them without the need to know the other; this is not the case for NIMD because $\pr(R=1|X, Y)$ and $ \pr(Y|X)$ are entangled together in the likelihood. Second, no identifiability issue is involved in MAR data problems, whereas models may not be identifiable based on NIMD even if fully parametric models are postulated on both $\pr(R=1|X, Y)$ and $\pr(Y|X)$ \citep{Heckman1979, Greenless1982, MDG2016}.

The identification issue must be overcome in NIMD problems before valid statistical inferences are made, otherwise the inference target is vague and meaningless. \cite{Robins1997} pointed out that it is impossible to identify the underlying parameters based on NIMD if both $\pr(R=1|X, Y)$ and $\pr(Y|X)$ are left completely unspecified. Attention has been paid to the case in which one of these probabilities is parametric or semiparametric and the other is left unspecified \citep{Tang2003, Qin2002, Chang2008, Kott2010, Kim2011, Wang2014, Riddles2016, Morikawa2017, Morikawa2016}. When no general identification results are available for NIMD, the joint distribution of the full data can be identified only under specific model assumptions. A popular and general condition for model identifiability with NIMD is the existence of an `instrumental variable' \citep{Wang2014} or equivalently an `ancillary variable' \citep{MT2016}, which does not affect the missingness probability but may affect the outcome regression function. \cite{Miao2019} found that the full data distribution can be identified with the aid of a shadow variable, which does not affect the missingness probability but may affect the observed outcome regression function $\pr(Y|X, R=1)$.

In recent years, many estimation approaches have been developed for identifiable model parameters based on NIMD \citep{Tang2018, Wang2021}. Of those, inverse probability weighting (IPW) methods are the ones used most commonly to deal with missing data including both MAR data \citep{Seaman2011} and NIMD under a parametric or semiparametric model for the missingness probability. Special cases include augmented IPW estimation or double robust estimation for MAR data \citep{Robins1995}, and the generalized method of moments with an instrumental variable for NIMD \citep{Wang2014, Zhao2015, Shao2016, Shao2018}. Double robust estimation under NIMD has also been investigated in the presence of an instrumental or ancillary variable \citep{Morikawa2016, MT2016, Ai2020, Liu2022}.

Playing a central role in IPW methods are unbiased IPW estimating equations. However, solving IPW equations numerically may involve non-convergence problems when the sample size is moderate and the missingness probability is high. In addition, these equations often have multiple roots, and it is challenging to identify the best root as a parameter estimate. Therefore, IPW methods may have low efficiency or even produce biased results. In this paper, under the most popular logistic regression (LR) model for the missingness probability, we find that the instability of IPW methods arises mainly from the fact that IPW equations involve the estimation of a moment-generating function (MGF) (Section~2), and such functions are notoriously unstable in general. As a remedy, we propose modelling the conditional outcome distribution given the covariates of the completely observed individuals by an accelerated time regression model or a semiparametric location-shift model. After transforming the model assumptions to a LR model for the missingness status of the outcome and covariate, we propose estimating the underlying parameters in the propensity score by maximizing a conditional likelihood (Section~3). Our estimation procedure circumvents the estimation of an MGF and hence overcomes the instability of IPW methods. 
%For convenience of presentation, all technical proofs are in the Appendices.
For clarity, all technical details are postponed to the  supplementary material.

 \section{Instability of inverse probability weighting methods}

We begin by discussing the instability of IPW estimation methods. Suppose that the missingness probability satisfies the commonly used LR model
\ba
\label{logistic}
\pr(R=1|x,y)
=
\pr(R=1|X=x,Y=y) = \frac{1}{1+\exp(\alpha_0 +x_1^\T \beta+y\gamma)},
\ea
where $x_1$ is equal to either $x$ or a subvector of $x$. Suppose that we have $n$ independent and identically distributed random vectors $(r_i, x_i, y_i)$ ($1\leq i\leq n$) from model \eqref{logistic}. In the absence of missing data, the score function of $(\alpha_0, \beta, \gamma)$ is
\bas
-\sum_{i=1}^n \left\{ r_i - \pr(R=1|x_i,y_i) \right\}
\times
(1,x_{i1}^\T,y_i)^\T.
\eas
Because $\pr(R=1|x_i,y_i)$ is uniformly bounded, the estimator of $(\alpha_0 , \beta, \gamma)$ that solves the score equations usually behaves stably.

In the presence of missing response (the $y_i$ with $r_i=0$ are missing), for any function $g(x)$, the IPW equation
\bas
\sum_{i=1}^n\left\{\frac{r_i}{\pr(R=1|x_i,y_i)}-1 \right\}g(x_i )=0
\eas
is equivalent to
\ba
\sum_{i=1}^n\left\{ r_i \exp(\alpha_0 +x_{i1}^\T \beta+y_i\gamma) +r_i -1 \right\}g(x_i )=0.
\label{gmm.ee}
\ea
Essentially, the first term involves estimating the MGF of $\pr(y|x,R=1)=\pr(Y=y|X=x,R=1)$. In the statistical literature, high-order moment estimates are notoriously unstable, let alone estimates of an MGF, and it is this unstable estimation that often causes the resulting IPW estimator to perform unstably. Moreover, a key estimating equation
\[
\sum_{i=1}^n\left\{ r_i \exp(\alpha_0 +x_{i1}^\T \beta+y_i\gamma) +r_i -1 \right\} =0,
\]
i.e.\ \eqref{gmm.ee} with $g(x)=1$, may have multiple roots for $\gamma$ even when $\alpha_0$ and $\beta$ are fixed to their true values.

\begin{figure}[h]
%\figurebox{4.2in}{}{}[root.eps]
\centerline{\includegraphics[scale=0.5]{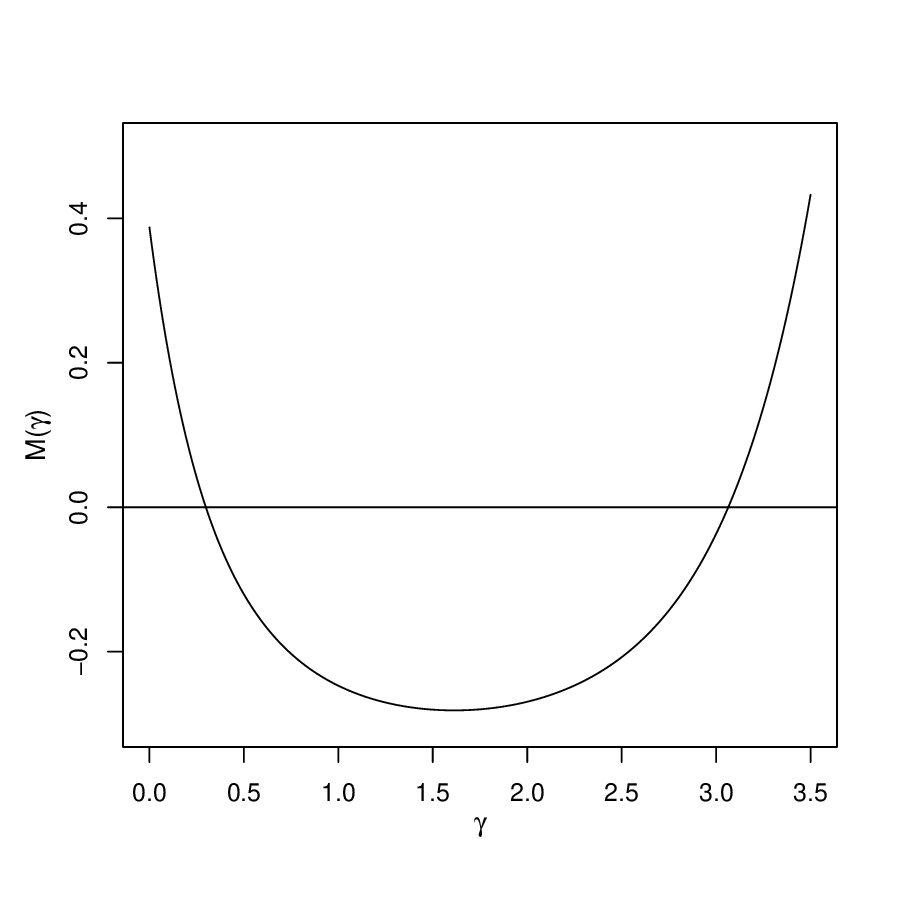}}
%\vspace{-0.2in}
\caption{\label{fig1}
Plot of $\gamma$ versus $M(\gamma)$ based on simulated data.
}
\end{figure}

For illustration, we generate a sample of size $n=20\,000$ from the following models:
\[
Y=X_{1} +X_2+\epsilon, ~~\pr(R=1|x,y)=\frac{1}{1+\exp(-1-x_1+3y)},
\]
where $X_1\sim N(0,2)$, $X_2\sim N(0,1)$, $\epsilon\sim N(0,1)$, and they are independent of each other. The proportion of missing data is around 33.6\%. We fix $(\alpha_0,\beta)$ to its true value $(-1,-1)$ and let
$$
M(\gamma)=
\frac{1}{n}\sum_{i=1}^nr_i\{ \exp(-1-x_{i1}+y_i\gamma) +1 \} -1.
$$
Figure~\ref{fig1} indicates clearly that $M(\gamma)$ has two roots. Next, we check the performance of the IPW approach, which solves the following three estimating equations
\[
\sum_{i=1}^n\left\{ r_i \exp(\alpha_0 +x_{i1} \beta+y_i\gamma) +r_i -1 \right\}
(1,x_{i1},x_{i2})^\T=0
\]
to estimate $(\alpha_0,\beta, \gamma)$. Even with a sample size as large as $n=20\,000$, based on 1000 repetitions, the means of the IPW estimates of $ \alpha_0$, $\beta$, and $ \gamma$ are $-2.86$, $-1.51$, and $4.16$, respectively, with standard deviations of $16.25$, $4.24$, and $7.08$, respectively. Clearly, the IPW method produces biased results and undesirably large standard deviations.

The above example shows that the IPW approach by assuming a parametric model for the propensity score function only may not be good enough to estimate the underlying parameters accurately. To overcome this issue, we can make a parametric model assumption, say $f(y|x, \xi)$, on $\pr(y|x, R=1)$ \citep{Lee2000, Riddles2016, Liu2022} in addition to the logistic model \eqref{logistic}. An obvious advantage of this model assumption over a completely parametric one for $\pr(y|x)$ is that it is checkable with available data. The instability of the IPW method can then be overcome by using the method due to \cite{Liu2022}.  A brief explanation follows, in which because the parameter $\xi$ can be estimated consistently by the maximizer of the conditional likelihood
\(
\prod_{i: r_i=1} f(y_i|x_i,\xi)
\)
under mild conditions, we take it as known for ease of presentation.

Model~\eqref{logistic} implies that
\ba
\label{drm.xy}
\pr(y, x |R=0) = \pr( y, x |R=1) \exp(\alpha_1 +x^\T_1 \beta+y\gamma),
\ea
where $\alpha_1 = \alpha_0 + \log\{\eta/(1-\eta)\}$ and $\eta = \pr(R=1)$. Therefore,
\ba
\pr(x |R=0)
&=&
\int \pr(y, x|R=0) dy \nonumber \\
&=&
\int \pr( y, x |R=1) \exp(\alpha_1 +x^\T_1 \beta+y\gamma) dy \nonumber\\
&=&
\pr(x |R=1) \exp\{\alpha_1 +x^\T_1 \beta + c(x; \gamma, \xi) \}, \label{drm0}
\ea
where $c(x; \gamma, \xi) = \log\{ E (e^{\gamma Y}|X=x,R=1) \}$. Using Bayes' formula,
we then have a new logistic model
\ba
\label{logistic-new}
\pr(R=1|x)
=
\pr(R=1|X=x) =
\frac{1}{1+\exp\{ \alpha_0 +x^\T_1 \beta + c(x; \gamma , \xi ) \}}.
\ea

The data $\{(x_i, r_i), i=1,2, \ldots, n\}$ are all completely observed and follow the new LR model \eqref{logistic-new}. Based on such data, the score equations for $(\alpha_0, \beta, \gamma)$ under model~\eqref{logistic-new} are
\ba
\label{liu-2022}
0&=& -\sum_{i=1}^n \left\{ r_i - \pr(R=1|x_i) \right\}
\Big(1, x_{i1}^\T, \nabla_{\gamma} c(x_i; \gamma, \xi) \Big)^\T,
\ea
where $\nabla_{\gamma} = \partial /\partial \gamma$. Again, because $\pr(R=1|x_i)$ is uniformly bounded, the resulting estimator of $(\alpha_0, \beta, \gamma)$ that solves the score equations behaves more stably than does the IPW estimator. In addition, \eqref{liu-2022} removes the effect of the randomness of the observed $y_i$ by conditioning on $x_i$. By the law of total variance, we conclude that the estimator determined by \eqref{liu-2022} has smaller asymptotic variance than that derived from the IPW equation.
These discussion also suggests  that a preferable approach to NIMD problems should make full use of the relationship between the distributions of $(X, R=0)$ and $(X, R=1)$.

\section{Semiparametric approach}

We assume the same LR model as in \eqref{logistic} for the missingness probability, but we relax the parametric model $f(y|x, \xi)$ for the conditional density function of $Y$ given $X=x$ and $R=1$ to a semiparametric location-scale model
\ba
y = \mu(x; \xi)+\epsilon,
\label{location-scale-model}
\ea
where $\mu(x; \xi)$ is a known function up to $\xi$ and $E(\epsilon)=0$. Here, the distribution of $\epsilon$, denoted as $f_\epsilon(\cdot)$, is completely unknown and $\epsilon$ is independent of $X$ given $R=1$. We wish to estimate $\tau = E(Y)$, and model~\eqref{location-scale-model} decreases to some extent the risk of model misspecification of a fully parametric model on $p(y|x,R=1)$.

\subsection{Parameter identifiability}

As we pointed out in the introduction, the issue of parameter identifiability always exists in NIMD problems regardless of which model assumptions are made. Before conducting valid statistical inference about the model parameters in models~\eqref{logistic} and \eqref{location-scale-model} and $\tau$, we need to investigate under what conditions they are identifiable.

Because the data with $R=1$ are all observed, we assume that $\xi$ and $f_\epsilon$ in model~\eqref{location-scale-model} are identifiable. Without loss of generality, we further regard $\xi$ and $f_\epsilon$ as known in this subsection. It remains to study the identifiability of $(\alpha_0, \beta, \gamma)$.

Let $M_1(t)=E(e^{t\epsilon})$ be the MGF of $\epsilon$. Under models~\eqref{logistic} and \eqref{location-scale-model}, $c(x; \gamma, \xi) =\gamma \mu(x;\xi)+\log M_1(\gamma)$, and the LR model \eqref{logistic-new} becomes
\ba
\label{logistic-new2}
\pi(x;\theta,\xi)=
\pr(R=1|x)=
\frac{1}{1+\exp\{ \alpha +x^\T_1 \beta + \gamma\mu(x;\xi)\}},
\ea
where $\alpha=\alpha_0+\log\{ M_1(\gamma)\}$ and $\theta=(\alpha,\beta^\T,\gamma)^\T$. Because $f_\epsilon$ is identifiable, the identifiability of $(\alpha_0, \beta, \gamma)$ is equivalent to that of $\theta$.

Because both $R$ and $X$ are observed, $\pr(R=1|x)$ is identifiable. With \eqref{logistic-new2},
$\theta$ is identifiable if and only if $\mu(x;\xi)$ is not a linear function of $x_1$. Let $x=(x_1^\T,x_2^\T)^\T$. If $x_2$ is not empty, then it is called an instrumental or shadow variable \citep{MT2016}. Here are two special cases: (i) if $\mu (x; \xi )$ is a nonlinear function of $x$, then $\theta$ is identifiable even if there is no instrumental variable or $x_2$ is empty; (ii) if $\mu (x; \xi )$ is a linear function of $x$ and $x_2$ is not empty, i.e.\ there exists an instrumental variable, then $\theta$ is identifiable.

\subsection{Estimation of model parameters and $\tau$}

In this subsection, we propose using a two-step procedure to estimate the model parameters $\xi$ and $\theta$ based on $\{(y_ir_i,x_i,r_i)\}_{i=1}^n$, where $y_i$ is observed if $r_i=1$ or missing otherwise.

In step~1, we estimate $\xi$ by using the least-squares method, i.e.
\begin{equation}
\label{est.hatxi}
\hat\xi=\arg\min_{\xi}\sum_{i=1}^n r_i\{y_i-\mu(x_i,\xi)\}^2.
\end{equation}
This step is implemented easily with the R function \texttt{lm} when $\mu (x; \xi )$ is a linear function of $x$ or \texttt{nls} when $\mu (x; \xi )$ is a nonlinear function of $x$. Note that the conditional likelihood of $\{r_i\}_{i=1}^n$ given $\{x_i\}_{i=1}^n$ is
$$
l_n(\theta,\xi)=
\sum_{i=1}^n\left[ r_i\log\{ \pi(x;\theta,\xi)\}
+(1-r_i) \log\{1- \pi(x;\theta,\xi)\}
\right].
$$

In step~2, we estimate $\theta$ by maximizing the conditional likelihood $l_n(\theta,\xi)$ with $\xi$ replaced by $\hat \xi$, i.e.
\ba
\label{est.hattheta}
\hat\theta
=\arg\max_{\theta}l_n(\theta,\hat\xi).
\ea
This step can be implemented with the R function \texttt{glm}. To construct an estimator of the response mean $\tau$, we rewrite it in terms of the underlying parameters $\eta$, $\xi$, and $\theta$. For any function $h(\cdot)$, it follows from models~\eqref{drm0} and \eqref{location-scale-model} that
\ba
E\{ h(Y)|R=0 \}
&=&
E [ h(Y) \exp\{\alpha_1 +X^\T_1 \beta + \gamma Y \}|R=1 ]
\label{key-equality1} \\
&=&
E [ h(Y) \exp\{\alpha_1 +X^\T_1 \beta + \gamma \mu(X, \xi ) \}|R=1 ] \cdot M_1(\gamma).
\label{key-equality2}
\ea
By the law of total expectation and \eqref{drm.xy}, we have
\bas
\tau
&=& E(Y|R=1)\pr(R=1)+ E(Y|R=0) \pr(R=0)\\
&=&\eta E\{\mu(X,\xi)|R=1\}+
(1-\eta) E\{ Y e^{\alpha_1+X^\T_1 \beta+Y\gamma} |R=1\}\\
&=&\eta E\{\mu(X,\xi)|R=1\}+
(1-\eta) E\{ \mu(X,\xi) e^{\alpha_1+X^\T_1 \beta+\gamma\mu(X,\xi)} |R=1\}M_1(\gamma)\\
&&+(1-\eta)E\{ e^{\alpha_1+X^\T_1 \beta+\gamma\mu(X,\xi)} |R=1\}M_2(\gamma)\\
&=&
E\{\mu(X,\xi)\}+(1-\eta)\frac{M_2(\gamma)}{M_1(\gamma)},
\label{mu.formula}
\eas
where $M_2(t)=E(\epsilon e^{t\epsilon})$ and we have used \eqref{key-equality1} in the second equality with $h(y)=y$ and used \eqref{key-equality2} in the last equality with $h(y)=1$.

Let $\hat\eta=\sum_{i=1}^nr_i/n$, $\hat\epsilon_i=y_i-\mu(x_i;\hat\xi)$, $\hat M_1(t)=\sum_{i=1}^n r_i e^{t\hat\epsilon_i} /\sum_{i=1}^n r_i$, and $\hat M_2(t)=\sum_{i=1}^n r_i\hat\epsilon_i e^{t\hat\epsilon_i} /\sum_{i=1}^n r_i$. Then a natural estimator of $\tau$ is
\ba
\hat \tau &=&
\frac{1}{n}\sum_{i=1}^n \mu(x_i,\hat\xi) +(1-\hat\eta) \frac{\hat M_2(\hat\gamma)}
{ \hat M_1(\hat\gamma)}.
\label{mle2}
\ea

\begin{remark}
The proposed estimator $\hat\tau$ relies on the regression model~\eqref{location-scale-model} and the missingness probability model~\eqref{logistic}. Based on the completely observed data $\{(y_ir_i,x_i,r_i)\}_{i=1}^n$, the correctness of \eqref{location-scale-model} can be verified by using the score test for non-constant error variance proposed by \cite{Breusch1979} and \cite{Cook1983}; this test is available in R as \texttt{ncvTest}. Because we do not have the observed values for $Y$ when $d_i=0$, model~\eqref{location-scale-model} cannot be verified directly; nevertheless, it can be transformed into the LR model~\eqref{logistic-new2}, and hence we can check \eqref{location-scale-model} indirectly by testing the goodness of fit of the LR model~\eqref{logistic-new2}. \cite{Hosmer1997} compared several goodness-of-fit tests for the LR model and recommended the one involving the unweighted sum of squares (USS) \citep{le1995}; this test is available as \texttt{residuals.lrm} in the R package \texttt{rms}.
\end{remark}

\subsection{Asymptotics}

In this subsection, we establish the asymptotic distributions of 
the proposed estimators $\hat \xi$, $\hat \theta$, and $\hat \tau$.  
Let $\eta_0$, $\xi_0$, $\theta_0$, and $\tau_0$ be the true values 
of $\eta$, $\xi$, $\theta$, and $\tau$, respectively, 
let $B^{\otimes 2}=BB^\T$ for any matrix or vector $B$, and 
let $\pi(x)=\pi(x;\theta_0,\xi_0)$.
 Define $\phi(x; \theta,\xi) = \alpha + x^\T_1 \beta + \gamma \mu(x ; \xi )$ 
 and $A_1 = E\left[R \left\{\nabla_{\xi} \mu(X; \xi_0) \right\}^{\otimes 2}\right]$, 
 $A_2 = E\left[ \pi(X)\{1-\pi(X)\} \left\{\nabla_{\theta} \phi(X; \theta_0,\xi_0) 
 \right\}^{\otimes 2} \right]$, $A_3 = E\left[ \pi(X)\{1-\pi(X)\}
 \left\{\nabla_{\theta} \phi(X; \theta_0,\xi_0) \right\} 
 \left\{\nabla_{\xi} \mu(X; \xi_0) \right\}^\T \right]$, 
 and $A_4 = E\left\{ \nabla_{\xi} \mu(X; \xi_0) \right\}$.  
Define 
$
H_1(x,y,r;\theta,\gamma)= r\exp[ \gamma\{ y- \mu(x;\xi) \} ]
$,
$
H_2(x,y,r;\theta,\gamma)=
r\{ y- \mu(x;\xi) \}\exp[ \gamma\{ y- \mu(x;\xi) \}] 
$,
and 
$S_1(x,y,r;\theta,\xi) = (S_{11}^\T(x,y,r;\theta), S_{12}^\T(x,y,r;\theta,\xi) )^\T$ with
\bas
S_{11}(x,y,r;\xi)
&=& r\{y-\mu(x;\xi)\}\{\nabla_{\xi} \mu(x; \xi)\}^\T,\\
S_{12}(x,y,r;\theta,\xi)
&=&\{r-\pi(x;\theta,\xi)\} \{ \nabla_{\theta} \phi(x; \theta_,\xi)\}^\T.
\eas 
\begin{assumption}
 There exists a function $g(x,y,r)$ with $\e\{g(X,Y,R)\}<\infty$ such that for $\theta$ and $\xi$ in a neighbourhood of $\theta_0$ and $\xi_0$,
\begin{description}
\item[(a)] the absolute value of each element of $ \nabla_{\xi\xi^\T}\mu(x, \xi)$ is bounded by $g(x,y,r)$;
\item[(b)] the absolute value of each element of the second derivatives of $S_1(x,y,r;\theta,\xi)$ with respect to $\theta$ and/or $\xi$ is bounded by $g(x,y,r)$;
\item[(c)] the absolute value of each element of the second derivatives of $H_1(x,y,r;\theta,\gamma)$ and $H_2(x,y,r;\theta,\gamma)$ with respect to $\theta$ and/or $\gamma$ is bounded by $g(x,y,r)$.
\end{description}
\end{assumption}

\begin{theorem}
\label{hat-xi-theta-asymptotic}
Suppose that $\mu(x;\xi)$ has a continuous second-order derivative with respect to $\xi$, 
  both $A_1$ and $A_2$ are positive definite and that Assumption 1 is satisfied. 
Assume that $(\theta_0,\xi_0)$ is the unique solution to $\e\{S_1(X,Y,R;\theta,\xi)\}=0$. 
As $n\rightarrow\infty$,
$
\sqrt{n}
( \hat \xi^\T - \xi_0^\T,
\hat \theta^\T - \theta_0^\T
)^\T
\rightarrow N(0, \Sigma)
$
in distribution,
where
$$
\Sigma
=\left(
\begin{array}{cc}
\sigma^2A_1^{-1}&-\gamma_0\sigma^2A_1^{-1}A_3^\T A_2^{-1}\\
-\gamma_0\sigma^2 A_2^{-1}A_3A_1^{-1}&
A_2^{-1}+\gamma^2_0\sigma^2A_2^{-1} A_3A_1^{-1}A_3^\T A_2^{-1}
\\
\end{array}
\right)
$$
and $\sigma^2=\var (\epsilon)$.
\end{theorem}

To use the results in Theorem~\ref{hat-xi-theta-asymptotic} to construct a Wald-type confidence interval (CI) for the parameters in $\xi$ or $\theta$, we need a consistent estimator for $\Sigma$, which can be constructed based on consistent estimators of $\gamma_0$, $\sigma^2$, and $A_1$--$A_3$. Reasonable estimators of $\gamma_0$, $\sigma^2$, and $A_1$ are $\hat\gamma$, $\hat\sigma^2=\sum_{i=1}^nr_i\epsilon_i^2/\sum_{i=1}^nr_i$, and
\ba
\label{hat.A1}
\hat A_1&=&n^{-1}\sum_{i=1}^n\left[r_i
\left\{
\nabla_{\xi} \mu(x_i; \hat\xi)
\right\}^{\otimes 2}
\right],
\ea
respectively. The estimators $\hat A_2$ and $\hat A_3$ for $A_2$ and $A_3$ can be constructed in a similar way to \eqref{hat.A1}. Inserting $\hat\sigma^2$, $\hat\xi$, and $\hat A_1$--$\hat A_3$ into $\Sigma$, we have an estimator $\hat\Sigma$ for $\Sigma$. With the results in Theorem~\ref{hat-xi-theta-asymptotic}, it can be verified that $\hat\Sigma$ is consistent with $\Sigma$.

To present the asymptotic distribution of $\hat\tau$, we need additional notation. Let $\mu_0 = \e\{\mu(X;\xi_0)\}$, $B_k = \e\left(R\epsilon^{k-1}e^{\gamma_0\epsilon}\right)$ for $k=1,2,3$, and $C_{k} = \e\{R\epsilon^{k-1} e^{\gamma_0\epsilon} \nabla_{\xi} \mu(X; \xi_0) \} $ for $k=1,2$. Furthermore, let $S=\left(S_0,S_1^\T,S_2^\T\right)^\T$ with $S_0 = R-\eta_0$, 
$S_1=S_1(X,Y,R;\theta_0,\xi_0)$,  and 
\bas 
S_2&=&\Big(\mu(X;\xi_0)-\mu_0,
Re^{\gamma_0\epsilon}-B_1,
R\epsilon e^{\gamma_0\epsilon}-B_2\Big) ^\T.
\eas
It can be verified that $\e(S)=0$. Denote $V=\var (S) = \e(S^{\otimes 2})$. Finally, let $p$ be the dimension of $\theta$, and let $e_p$ be a $p\times1$ vector with the $p$th element being 1 and the other elements being 0.

\begin{theorem}
\label{hat-tau-normality}
Suppose that the conditions in Theorem \ref{hat-xi-theta-asymptotic} are satisfied. 
As $n\rightarrow\infty$, $\sqrt{n} (\hat \tau - \tau_0) \to N(0,\sigma^2_{\tau})$ in distribution, where $\sigma^2_\tau=D^\T V D$. Here,
$$
D=\left(-\frac{B_2}{B_1},H_1,H_2,1,-(1-\eta_0)\frac{B_2}{B_1^2},\frac{1-\eta_0}{ B_1 }\right)^\T
$$
with
\bas
H_1&=&A_4^\T A_1^{-1}+(1-\eta_0)\left(
\frac{B_2}{B_1^2}C_1^\T-\frac{1}{B_1}C_1^\T-\frac{\gamma_0}{B_1}C_2^\T\right)A_1^{-1}\\
&&+ \frac{(1-\eta_0)\gamma_0}{B_1^2}
(B_2 -B_1B_3)
e_p^\T A_2^{-1} A_3 A_1^{-1}
\eas
and
$
H_2 =
(B_2^2 -B_1B_3)e_p^\T A_2^{-1} (1-\eta_0)/B_1^2.
$
\end{theorem}

To construct a Wald CI for $\tau$ based on Theorem~\ref{hat-tau-normality}, we need a consistent estimator of $\sigma^2_\tau$, which depends on $V$. We first construct a desirable estimator for $V$. Let $\hat\eta=n^{-1}\sum_{i=1}^n r_i$, $\hat\mu_0=n^{-1}\sum_{i=1}^n \mu(x_i; \hat\xi)$, $\hat B_k=n^{-1}\sum_{i=1}^n\left(r_i\hat\epsilon^{k-1}_ie^{\hat\gamma\hat\epsilon_i}\right)$, and $\hat S_i=\left(\hat S_{0i},\hat S_{1i}^\T,\hat S_{2i}^\T\right)^\T$ with $\hat S_{0i} = r_i-\hat \eta$,
\bas
\hat S_{1i}&=&\left(r_i\hat\epsilon_i\{\nabla_{\xi} \mu(x_i; \hat\xi)\}^\T,
\{r_i -\pi(x_i;\hat\theta,\hat\xi)\} \{ \nabla_{\theta} \phi(x_i; \hat\theta,\hat\xi)\}^\T
\right)^\T,\\
\hat S_{2i}&=&\left(\mu(x_i;\hat \xi)-\hat \mu_0,
r_ie^{\hat \gamma\hat \epsilon_i}-\hat B_1,
r_i\hat \epsilon_i e^{\hat \gamma \hat \epsilon_i}-\hat B_2\right) ^\T.
\eas
Then a natural estimator for $V$ is $\hat V=n^{-1}\sum_{i=1}^n \hat S_i\hat S_i^\T$. With the results in Theorem~\ref{hat-xi-theta-asymptotic}, it can be verified that $\hat V$ is consistent with $V$. Using the techniques used to construct $\hat\Sigma$, we can construct a consistent estimator $\hat D$ for $D$. Finally, we estimate $\sigma^2_\tau$ by $\hat\sigma^2_\tau=\hat D^\T \hat V \hat D$, and a $100(1-a)\%$ CI of $\tau$ is
$
\mathcal{I}_{{\tau}}=[\hat\tau-Z_{1-a/2}\hat\sigma_\tau, \hat\tau+Z_{1-a/2}\hat\sigma_\tau],
$
where $Z_{1-a/2}$ is the $(1-a/2)$th quantile of $N(0,1)$.

\section{Simulation}
\subsection{Setup}

We compare the proposed estimator $\hat\tau $ of $\tau$ with the following competitors developed recently in the literature:
\begin{itemize}
\item $\hat\tau_{\footnotesize P}$, the maximum empirical likelihood estimator due to \cite{Liu2022}, where the model for $Y$ given $X=x$ and $R=1$ is \eqref{location-scale-model} with the error distribution being normal;
\item $\hat\tau_{\footnotesize A1}$, the parametric adaptive method due to \cite{Morikawa2016}, where the working model for $Y$ given $X=x$ and $R=1$ is \eqref{location-scale-model} with the error distribution being normal;
\item $\hat\tau_{\footnotesize A2}$, the nonparametric adaptive method due to \cite{Morikawa2016}, where the working model for $Y$ given $R=1$ and $X$ is fully nonparametric and the kernel function and bandwidth are those recommended by \cite{Morikawa2016};
\item $\hat\tau_{\footnotesize Gk}$, the generalized moment method due to \cite{Ai2020}, where the basis functions comprise $\left\{\prod x_{1}^{i_1}\cdots x_d^{i_d}: i_1\geq0,\ldots,i_d\geq0,\sum_{j=1}^d i_j\leq k \right\}$.
\end{itemize}
We generate data from two examples.

\begin{example}
\label{exm1}
Suppose that there are only two covariates $X_1\sim N(1,1)$ and $X_2\sim N(0,1)$, which are independent. We set $\pr(R=1|x, y) = 1/\{1+\exp(\alpha_0-0.4x_1 +0.5y)\}$ and $y=2.5-x_1+1.5x_2+\epsilon $ given $X=x$ and $R=1$. We consider two values of $\alpha_0$, i.e.\ $-1.7$ and $-1.2$, and the missingness probability increases as $\alpha_0$ increases.
\end{example}

\begin{example}
\label{exm2}
Suppose that there is only one covariate $X\sim N(0,1)$. We set $\pr(R=1|x, y) = 1/\{1+\exp(\alpha_0-0.4x +0.5y)\}$ and $y=2-x+x^2+\epsilon $ given $X=x$ and $R=1$. We consider two values of $\alpha_0$, i.e.\ $-2.7$ and $-2.2$, and again the missingness probability increases as $\alpha_0$ increases.
\end{example}

In both examples, we consider two distributions for $\epsilon$, i.e.\ $2/3N(-\delta,4-3\delta^2)+1/3N(2\delta,4)$ with $\delta=0$ and $1$. The error distribution is just $N(0,4)$ when $\delta=0$, and it is a normal mixture $2/3N(-1,1)+1/3N(2,4)$ when $\delta=1$. The true values of $\tau$ and the missingness probability $\pr(R=0)$ for the two examples are tabulated in Table~\ref{truemu}.

\begin{table}[t]
\caption{True values of $\tau$ and missingness probability $\pr(R=0)$ in examples~\ref{exm1} and \ref{exm2}.
\label{truemu}
}
\centering
\tabcolsep 5pt
\renewcommand{\arraystretch}{1}
\begin{tabular}{ccccc ccccc}
\hline
Example & $\alpha_0$ & $\delta$ & $\tau$ & $P(R=0)$ & Example & $\alpha_0$ & $\delta$ & $\tau$ & $P(R=0)$\\
\hline
1 & $-1.7$ & 0 & 2.177 & 0.339 & 1 & $-1.7$ & 1 & 2.587 & 0.369\\
1 & $-1.2$ & 0 & 2.364 & 0.432 & 1 & $-1.2$ & 1 & 2.868 & 0.465\\
2 & $-2.7$ & 0 & 3.677 & 0.338 & 2 & $-2.7$ & 1 & 4.088 & 0.369\\
2 & $-2.2$ & 0 & 3.869 & 0.434 & 2 & $-2.2$ & 1 & 4.381 & 0.469\\
\hline
\end{tabular}

\end{table}

\subsection{Results for point estimates}

We summarize the results in terms of  relative bias (RB) and mean square error (MSE) for estimating $\tau$ in Tables~\ref{RB.MSE.example1} and \ref{RB.MSE.example2}. Note that we encountered numerical problems in implementing the adaptive methods due to \cite{Morikawa2016} and the generalized moment method due to \cite{Ai2020}. In the simulation study, we count the number of non-convergence or non-reliable cases for each method, where a non-reliable case is one simulation repetition in which either the estimate of $\tau$ is outside the range $[-10,10]$ or the estimate of $\gamma$ is outside the range $[-3,3]$. For each method, the RB and MSE results reported are evaluated based on the convergent cases.

From the simulation results, we make the following observations. 1) The proposed estimator $\hat \tau$ performs similarly to the estimator $\hat \tau_P$ due to \cite{Liu2022} when the error distribution is normal, while $\hat\tau$ has smaller RB and MSE than those of $\hat \tau_P$ when the error distribution is a mixture of normal distributions. This shows the robustness of the proposed method. 2) As we discussed before, the methods due to \cite{Morikawa2016} and \cite{Ai2020} may experience numerical issues, which become more prominent when the sample size is small, the missingness probability is high, and/or no instrumental variable exists. 3) The parametric adaptive method due to \cite{Morikawa2016} produces much larger MSE than does our estimator $\hat \tau$, although the former far outperforms their nonparametric adaptive method in terms of both RB and MSE. 4) The performance of the generalized moment method due to \cite{Ai2020} depends on the choice of basis functions. When the number of basis functions increases, the bias increases and the variance decreases. Compared with the proposed method, the generalized moment method usually produces an MSE that is at least 50\% and can be double or ever higher.

\begin{table}[!http]
\begin{center}
\renewcommand{\arraystretch}{0.6}
\caption{
Relative bias (RB; $\times 100$), mean square error (MSE; $\times 100$), and number of non-convergence or non-reliable cases (NCR) of six estimators of $\tau$ (example~\ref{exm1}).
\label{RB.MSE.example1}
}
\tabcolsep 2pt
\begin{tabular}{l rr rr rr rr rr rr rr rr rr rr rr rr }\hline
&RB&MSE&NCR&RB&MSE&NCR&RB&MSE&NCR&RB&MSE&NCR\\\hline
&\multicolumn{12}{c }{$\epsilon \sim N(0,4)$}\\
&\multicolumn{6}{c }{$\alpha_0=-1.7$} & \multicolumn{6}{c }{$\alpha_0=-1.2$}\\
&\multicolumn{3}{c}{$n=500$}&\multicolumn{3}{c}{$n=2000$}&\multicolumn{3}{c}{$n=500$}&\multicolumn{3}{c}{$n=2000$}\\\hline

$\hat\tau $ &$-0.16$&4.18&0.00&0.17&1.06&0.00&$-0.26$&5.64&0.00&0.04&1.39&0.00\\
$\hat\tau_{P}$ &0.02&3.95&0.00&0.18&0.97&0.00&0.00&5.03&0.00&0.09&1.23&0.00\\
$\hat\tau_{A1}$ &$-1.12$&11.57&13.00&0.09&1.21&1.00&$-1.77$&21.45&14.00&$-0.11$&1.72&0.00\\
$\hat\tau_{A2}$ &$-30.39$&46.39&0.00&$-29.96$&43.16&0.00&$-35.88$&75.02&0.00&$-35.20$&69.99&0.00\\
$\hat\tau_{G1}$&1.70&7.85&13.00&0.95&2.28&4.00&2.00&11.04&16.00&1.24&4.55&5.00\\
$\hat\tau_{G2}$&$-6.08$&6.22&6.00&$-2.15$&1.45&0.00&$-8.02$&9.74&10.00&$-2.93$&2.21&0.00\\\hline
&\multicolumn{12}{c }{$\epsilon \sim 2/3N(-1,1)+1/3N(2,4)$}\\
&\multicolumn{6}{c }{$\alpha_0=-1.7$} & \multicolumn{6}{c }{$\alpha_0=-1.2$}\\
&\multicolumn{3}{c}{$n=500$}&\multicolumn{3}{c}{$n=2000$}&\multicolumn{3}{c}{$n=500$}&\multicolumn{3}{c}{$n=2000$}\\\hline

$\hat\tau $ &$-0.27$&10.09&0.00&$-0.10$&2.38&0.00&$-0.67$&14.49&0.00&$-0.26$&3.60&0.00\\
$\hat\tau_{P}$ &$-13.18$&17.17&0.00&$-13.39$&13.31&0.00&$-15.20$&26.54&0.00&$-15.27$&20.96&0.00\\
$\hat\tau_{A1}$ &$-0.55$&32.81&31.00&$-0.22$&7.89&4.00&$-1.29$&44.22&41.00&$-0.37$&7.55&7.00\\
$\hat\tau_{A2}$ &$-41.59$&118.50&0.00&$-41.37$&115.20&0.00&$-47.39$&188.29&0.00&$-46.95$&182.17&0.00\\
$\hat\tau_{G1}$ &2.39&18.50&49.00&2.62&9.53&14.00&2.46&29.47&57.00&2.97&14.20&13.00\\
$\hat\tau_{G2}$ &$-11.83$&19.34&12.00&$-5.04$&5.17&0.00&$-14.93$&33.29&13.00&$-6.47$&8.62&0.00\\\hline
\end{tabular}
\end{center}
\end{table}

\begin{table}[!http]
\begin{center}
\renewcommand{\arraystretch}{1}
\caption{
Relative bias (RB; $\times 100$), mean square error (MSE; $\times 100$), and number of non-convergence or non-reliable cases (NCR) of seven estimators of $\tau$ (example~\ref{exm2}).
\label{RB.MSE.example2}
}
\tabcolsep 2pt

\begin{tabular}{l rr rr rr rr rr rr rr rr rr rr rr rr }\hline
&RB&MSE&NC&RB&MSE&NC&RB&MSE&NC&RB&MSE&NC\\\hline
&\multicolumn{12}{c }{$\epsilon \sim N(0,4)$}\\
&\multicolumn{6}{c }{$\alpha_0=-2.7$} & \multicolumn{6}{c }{$\alpha_0=-2.2$}\\
&\multicolumn{3}{c}{$n=500$}&\multicolumn{3}{c}{$n=2000$}&\multicolumn{3}{c}{$n=500$}&\multicolumn{3}{c}{$n=2000$}\\\hline
$\hat\tau $&0.25&4.99&0.00&0.14&1.15&0.00&0.36&7.14&0.00&0.21&1.68&0.00\\
$\hat\tau_{P}$ &0.39&4.59&0.00&0.20&1.06&0.00&0.55&6.60&0.00&0.27&1.52&0.00\\
$\hat\tau_{A1}$&$-1.01$&61.34&133.00&$-0.03$&2.64&4.00&$-0.87$&107.69&195.00&0.06&7.05&10.00\\
$\hat\tau_{A2}$&$-11.81$&22.72&16.00&$-8.71$&11.30&3.00&$-14.77$&38.04&18.00&$-10.73$&18.76&2.00\\
$\hat\tau_{G2}$&4.25&14.48&79.00&2.83&5.65&3.00&5.33&23.20&101.00&4.17&11.44&1.00\\
$\hat\tau_{G3}$&$-3.99$&15.57&144.00&$-1.48$&2.84&10.00&$-4.06$&20.79&135.00&$-1.54$&3.96&16.00\\
$\hat\tau_{G4}$&$-6.96$&17.07&92.00&$-3.93$&4.38&16.00&$-7.58$&24.74&59.00&$-4.61$&6.69&5.00\\\hline
&\multicolumn{12}{c }{$\epsilon \sim 2/3N(-1,1)+1/3N(2,4)$}\\
&\multicolumn{6}{c }{$\alpha_0=-2.7$} & \multicolumn{6}{c }{$\alpha_0=-2.2$}\\
&\multicolumn{3}{c}{$n=500$}&\multicolumn{3}{c}{$n=2000$}&\multicolumn{3}{c}{$n=500$}&\multicolumn{3}{c}{$n=2000$}\\\hline
$\hat\tau $&$-0.27$&11.94&0.00&0.15&3.05&0.00&$-0.10$&19.15&0.00&0.20&5.18&0.00\\
$\hat\tau_{P}$ &$-8.32$&17.87&0.00&$-8.35$&13.16&0.00&$-9.68$&27.27&0.00&$-9.89$&21.00&0.00\\
$\hat\tau_{A1}$&$-0.70$&180.78&208.00&0.24&21.22&29.00&0.98&428.26&575.00&0.76&39.74&52.00\\
$\hat\tau_{A2}$&$-20.57$&75.86&100.00&$-16.35$&46.33&32.00&$-24.77$&124.82&209.00&$-19.90$&78.37&32.00\\
$\hat\tau_{G2}$&4.18&26.72&164.00&6.28&22.98&16.00&3.89&42.10&213.00&8.00&41.48&38.00\\
$\hat\tau_{G3}$&$-6.15$&27.81&149.00&$-1.96$&8.77&22.00&$-7.40$&44.09&139.00&$-2.34$&17.44&40.00\\
$\hat\tau_{G4}$&$-9.22$&30.05&61.00&$-5.33$&12.07&12.00&$-11.37$&52.25&41.00&$-6.60$&23.75&16.00\\\hline
\end{tabular}
\end{center}
\end{table}

\subsection{Results for confidence intervals}

In this subsection, we evaluate the performance of the proposed CI $\mathcal{I}_{{\tau}}$ for $\tau$. To improve the performance of $\mathcal{I}_{{\tau}}$, we also consider the bootstrap $t$-type CI, which is $\mathcal{I}_{{\tau}}$ with the normal quantile replaced by the corresponding nonparametric bootstrap quantiles based on 1000 bootstrap samples; we denote the bootstrap $t$-type CI by $\mathcal{I}_{{\tau}}^B$. The simulated coverage probabilities of $\mathcal{I}_{{\tau}}$ and $\mathcal{I}_{{\tau}}^B$ are provided in Table~\ref{CIs}. We did not compare the proposed CIs with those based on other methods because $\hat\tau_P$ produces biased results when the error distribution is not normal, and the methods due to \cite{Morikawa2016} and \cite{Ai2020} may experience numerical issues.

Table~\ref{CIs} shows that $\mathcal{I}_{{\tau}}$ has accurate coverage probabilities when the error distribution is either normal or non-normal with $n=2000$, but it experiences under-coverage when the error distribution is non-normal with $n=500$. By contrast, the bootstrap $t$-type CI has accurate coverage probabilities in all situations and so is recommended in applications.

\begin{table}[!ht]
\renewcommand{\arraystretch}{1}
\caption{
Coverage probabilities of $\mathcal{I}_{{\tau}}$ and $\mathcal{I}_{{\tau}}^B$ at 95\% nominal level.
\label{CIs}
}
\tabcolsep 10pt
\begin{center}
\begin{tabular}{c c c c c c }\hline
Example&$\alpha_0$& $\mathcal{I}_{{\tau}}$ &$\mathcal{I}_{{\tau}}^B$& $\mathcal{I}_{{\tau}}$ &$\mathcal{I}_{{\tau}}^B$ \\\hline
&\multicolumn{5}{c }{$\epsilon \sim N(0,4)$}\\
&&\multicolumn{2}{c }{$n=500$} & \multicolumn{2}{c }{$n=2000$}\\\hline
1&$-1.7$&95.0&94.7&95.4&95.0\\
1&$-1.2$&94.1&94.8&95.4&94.7\\
2&$-2.7$&95.0&95.0&95.5&94.7\\
2&$-2.2$&94.8&95.4&95.6&94.8\\\hline
&\multicolumn{5}{c }{$\epsilon \sim 2/3N(-1,1)+1/3N(2,4)$}\\
&&\multicolumn{2}{c }{$n=500$} & \multicolumn{2}{c }{$n=2000$}\\\hline
1&$-1.7$&93.2&94.4&95.0&95.2\\
1&$-1.2$&92.6&94.4&94.7&95.0\\
2&$-2.7$&92.7&94.7&95.0&95.0\\
2&$-2.2$&92.6&95.0&94.7&95.4\\\hline
\end{tabular}
\end{center}
\end{table}

\section{Real examples}

We analyze two real examples for illustration. The first example involves human immunodeficiency virus (HIV) data from the AIDS Clinical Trials Group Protocol 175 (ACTG175) \citep{Hammer1996}, in which $n=2139$ HIV-infected patients were enrolled. The patients were divided randomly into four arms according to the regimen of treatment that they received: (I) zidovudine monotherapy, (II) zidovudine + didanosine, (III) zidovudine + zalcitabine, and (IV) didanosine monotherapy. For illustration, we consider the patients in arm~III; analysis for the other arms can be conducted similarly.

The data record many measurements from each patient, including their age (in years), weight (in kilograms), CD4 cell count at baseline (cd40), CD4 cell count at $20 \pm 5$ weeks (cd420), CD4 cell count at $96 \pm 5$ weeks (cd496), CD8 cell count at baseline (cd80), CD8 cell count at $20 \pm 5$ weeks (cd820), and arm number (arms). The data are available from the R package \texttt{speff2trial}. The effectiveness of an HIV treatment can be assessed by monitoring the CD4 cell counts of HIV-positive patients: an increase indicates an improvement in the patients' health. An interesting problem is determining the mean of the CD4 cell counts after the patients were treated for about 96 weeks. We take cd496 as the response variable $Y$, and we take age, weight, cd40, cd420, cd80, and cd820 as covariates $X_1, \ldots, X_6$, respectively. Because of either the trial ending or lack of follow-up, 35.7\% of the patients' responses were missing.

We take $X = (X_1,X_3, X_4, X_6)$ and consider the following location-scale model for $Y$ given $X=x$ and $R=1$:
\ba
\label{real.exm1.reg}
y=\mu(x, \xi) +\epsilon
=\xi_1 + \xi_2 x_1
+
\xi_3 x_3+\xi_4 x_4
+\xi_5x_6+
+ \xi_6 x_1^2+ \xi_7 x_4^2+\epsilon.
\ea
This model is chosen by the all-subset selection method coupled with the Akaike information criterion among the six covariates and their quadratic terms. The score test for non-constant error variance proposed by \cite{Breusch1979} and \cite{Cook1983} gives a $p$-value of around 0.560, which supports the assumption of constant error variance. To check whether the normality assumption is suitable for the errors in \eqref{real.exm1.reg}, we perform a Shapiro--Wilk test for the residuals, which produces a $p$-value of around $4.28 \times 10^{-8}$. Therefore, we have no evidence against the location-scale mode with a constant error in \eqref{real.exm1.reg}, but the normal error assumption may not be suitable.

Next, we consider modelling the missingness probability. Recall that the location-scale model \eqref{location-scale-model} and the missingness probability model \eqref{logistic} imply the LR model \eqref{logistic-new2} for $R$ given $X=x$. With the available data $\{(x_i,r_i)\}_{i=1}^n$ and the chosen model $\mu(x;\xi)$, we use the all-subset section method coupled with the Akaike information criterion to choose the most appropriate model for $R$ given $X=x$, which is
\ba\label{real.exm1.miss1}
\pr(R=1|x)=
\frac{1}{
1+\exp\{\alpha_0+ x_1 \beta_1 + x_3\beta_2 + \gamma\mu(x;\xi) \} }.
\ea
To evaluate the goodness of fit of this model, the USS test due to \cite{le1995} gives a $p$-value of around 0.722. Therefore, model~\eqref{real.exm1.miss1} provides a reasonable fit for $R$ given $X=x$, which together with the location-scale model~\eqref{real.exm1.reg} for $Y$ given $X=x$ and $R=1$ implies that the missingness probability model
\ba\label{real.exm1.miss2}
\pr(D=1|x,y)=
\frac{1}{
1+\exp(\alpha_0+ x_1 \beta_1 + x_3\beta_2 + y \gamma)}
\ea
is suitable.

We now apply the proposed estimator $\hat\tau$ and CI $\mathcal{I}_{{\tau}}^B$, based on models \eqref{real.exm1.reg} and \eqref{real.exm1.miss2}, to the ACTG175 data for patients in arm~III. For comparison, we include $\hat\tau_P$, $\hat\tau_{A1}$, and $\hat\tau_{G1}$ and their corresponding bootstrap percentile CIs based on 1000 bootstrap samples. We do not include the nonparametric adaptive method due to \cite{Morikawa2016} and $\hat \tau_{Gk}$ due to \cite{Ai2020} for $k\geq 2$ because both methods show large biases in our simulation study. The results are summarized in Table~\ref{tab-real}. The four point estimates $\hat\tau$, $\hat\tau_P$, $\hat\tau_{A1}$, and $\hat\tau_{G1}$ are slightly different. Based on our simulation results for the non-normal case, $\hat\tau$ always gives the smallest bias, so we reason that the result for $\hat\tau$ is more reliable. The CI based on $\hat\tau$ is the smallest length among the four methods, which shows the advantage of the proposed method for the non-normal error case.

\begin{table}[!ht]
\caption{
Analysis results  for ACTG175 data ($Y=$ CD496 cell counts  of patients in arm~III)
and PPTV data ($Y=$ logarithm of difference in PPVT score between 1986 and 1992 of children).
\label{tab-real} }
\vspace{0.1in}
\tabcolsep 4pt
\renewcommand{\arraystretch}{1}\centering
\begin{tabular}{  cccc}\hline
  & Point estimate & Interval estimate & Length of CI \\\hline
\multicolumn{4}{c}{ ACTG175 data}  \\
$\hat \mu $ &308.98&[279.68, 330.97] &51.29 \\
$\hat \mu_{P}$ &305.72&[275.63, 333.07] &57.44 \\
$\tilde \mu_{A1}$ &310.39 &[282.87, 340.55] &57.68 \\
$\tilde \mu_{G1}$ &313.68 &[286.36, 339.16] & 52.80\\\hline
 \hline
\multicolumn{4}{c}{ PPVT data}  \\ 
$\hat \tau $ &4.302&[4.278, 4.323] &0.045 \\
$\hat \tau_{P}$ & 4.303&[4.275,4.329] &0.054 \\
$\hat \tau_{A1}$ &4.301 &[4.273, 4.325] &0.052 \\
$\hat \tau_{G1}$ &4.297&[4.255, 4.325] &0.069\\\hline
\end{tabular}
\end{table}

The second example involves the Peabody Picture Vocabulary Test (PPVT) data analyzed by \cite{Chen2021}, which were collected as part of the National Longitudinal Survey of Youth (NLSY79 Child and Young Adult cohort). The PPVT comprises a number of items, each of which involves four pictures; the interviewer says a word out loud, and the child selects the picture of the four that best describes the word's meaning. The data come from test results between 1986 and 1992 for children who were aged between three and four years at the 1986 assessment and whose mothers reported nonzero income in at least one year between 1986 and 1992. In total, $n=557$ children are in the sample.

We let the response $Y$ be the logarithm of the difference in PPVT score between 1986 and 1992. Following \cite{Chen2021}, we consider seven covariates: gender (1 = Male, 0 = Female; $x_1$), race (1 = White, 0 = Other; $x_2$), mother's hourly income ($x_3$), mother's education (1 = $>12$ years, 0 = $\leq 12$ years; $x_4$), and three dummy variables ($x_5$--$x_7$) that classify the data by the four quartiles of the 1986 PPVT score, i.e.\ $(x_5,x_6,x_7) = (0,0,0)$, $(1,0,0)$, $(0,1,0)$, or $(0,0,1)$ if a 1986 PPVT score is in the first, second, third, or forth quartile, respectively. For various reasons, such as motivation, family influence, and perceived poor performance, only 387 valid assessments were obtained in 1992, which gives a missing-data rate of 30.5\%.

We take $X = (X_1,\ldots,X_7)$ and consider the following location-scale model for $Y$ given $X=x$ and $R=1$:
\ba
\label{real.exm2.reg}
y=\mu(x, \xi) +\epsilon
=\xi_1 + \xi_2 x_1+\cdots
+\xi_8 x_7+\epsilon.
\ea
The $p$-value of the score test for non-constant error variance is around 0.296, and that for the Shapiro--Wilk test on residuals is around $5.74 \times 10^{-12}$. These results indicate that the location-scale model~\eqref{real.exm2.reg} is reasonable for $Y$ given $X=x$ and $R=1$ but that the normality assumption for the error may not be suitable.

Following \cite{Chen2021}, we use $x_5$--$x_7$ as instrumental variables and consider the following
missingness probability model:
\ba\label{real.exm2.miss2}
\pr(R=1|x,y)=
\frac{1}{
1+\exp(\alpha_0+ x_1 \beta_1 +\cdots+\beta_4x_4 + y \gamma)}.
\ea
The USS test due to \cite{le1995} for the goodness of fit of the induced LR model
\ba\label{real.exm2.miss1}
\pr(R=1|x)=
\frac{1}{
1+\exp\{\alpha+ x_1 \beta_1 +\cdots+\beta_4x_4 + \gamma\mu(x;\xi)\}}.
\ea
gives a $p$-value of around 0.737, which supports the missingness probability model~\eqref{real.exm2.miss2}.

We now apply the proposed estimator $\hat\tau$ and CI $\mathcal{I}_{{\tau}}^B$, based on \eqref{real.exm2.reg} and \eqref{real.exm2.miss2}, to the PPVT data. For comparison, we also include the results for $\hat\tau_{P}$, $\hat\tau_{A1}$, and $\hat\tau_{G1}$. The results are summarized in Table~\ref{tab-real}. For these data, all four methods give similar point estimates, but the proposed CI $\mathcal{I}_{{\tau}}^B$ has the shortest length, which again shows the advantage of the proposed method for non-normal data.

\section{Concluding remarks}

As mentioned in the beginning of the introduction, inference for problems involving NIMD is much harder than that for those involving MAR data. In general, one must either specify the propensity score or make a parametric assumption about the outcome given covariates, otherwise the underlying models are not identifiable. The existing literature shows that it is possible to identify the propensity-score parameters if one specifies a parametric model for the propensity score only and leaves the regression model arbitrary. However, the IPW estimating equations involve an MGF, and such functions can suffer from slow convergence; consequently, such equations are very unstable numerically and may have multiple roots. Our extensive simulation studies have shown that inference based on a parametric logistic propensity-score model alone is very unreliable, and one must make some assumption about the regression model, either parametrically or semiparametrically. In this paper, we have proposed an innovative method for NIMD problems when a parametric model for the propensity score is specified and the observed outcome follows a location-shift model with an unspecified error distribution. In the proposed estimation approach, we focus on the newly developed LR model for the observed covariates. The resultant estimating equations for the propensity-score parameters are bounded, which makes them converge much faster. Extensive simulations have shown that our method far outperforms the existing ones.

\end{document}